\documentclass[10pt]{article}
\usepackage[margin=2cm]{geometry}

\usepackage{hyperref,graphicx,amsmath,amssymb}

\begin{document}


\title{Simulating urban dynamics and international governance of transportation infrastructure projects}

\author{Juste Raimbault$^{1,2,3,\ast}$\medskip\\
$^{1}$ Center for Advanced Spatial Analysis, University College London\\
$^{2}$ UPS CNRS 3611 ISC-PIF\\
$^{3}$ UMR CNRS 8504 G{\'e}ographie-cit{\'e}s\medskip\\
$^{\ast}$ \texttt{juste.raimbault@polytechnique.edu}
}
\date{}



\maketitle

\begin{abstract}
Systems of cities at the macroscopic scale have their trajectories conditioned by the evolution of infrastructure networks. This leads to complex planning and management situations in the particular case of international transportation infrastructure projects. To understand such dynamics and anticipate future sustainable trajectories, we introduce a co-evolution model between cities and transportation networks at the international scale, which simulates network growth by including transportation governance. The model is applied to synthetic systems of cities and the parameter space systematically explored, showing strong interactions between urban dynamics and governance structure. We also study optimisation patterns as compromises between construction cost and accessibility gain, with possible future applications to sustainable long-term planning of international transport projects.
\end{abstract}

\section{Introduction}

Interactions between transportation networks and territories shape urban dynamics on multiple spatial and temporal scales. The evolution of land-use is highly dependent on the spatial distribution of accessibility \cite{wegener2004land}, although it remains difficult to anticipate the long terms effects of a rapid evolution of transportation infrastructure \cite{raimbault:halshs-02265423}. The modeling and simulation of complex urban systems is an approach to understand such entangled processes. \cite{pumain2018evolutionary} describes an evolutionary theory of systems of cities, which suggests in particular that cities co-evolve in a complex network of interactions. In that context, \cite{pumain2017urban} recall the main stylised facts that such a theory must encompass, introduce simulation models to reproduce these and provide tools and methods to extract knowledge from simulation models, mainly included within the OpenMOLE open-source platform \cite{reuillon2013openmole}.

Cities and transportation networks can thus be characterised as co-evolving in several cases \cite{raimbault2018caracterisation}. At the macroscopic scale, the support of urban interactions are links of the transport network, which dynamics are ruled by transport investments which depend partly on the trajectory of cities. An urban dynamic model including network effects was calibrated for France by \cite{raimbault2020indirect}, and extended into a co-evolution model by \cite{raimbault2021modeling}. While the influence of access or centrality on population dynamics is relatively easy to capture with spatial interaction models, general rules for the growth of spatial transportation networks remain an open question \cite{barthelemy2018morphogenesis}. To take into account political decisions and governance processes into network growth models, \cite{raimbault2021introducing} propose a co-evolution model at the scale of city regions which combines a land-use transport interaction model with an agent-based models of planning actors taking investment decisions to build transport links.

Large infrastructure projects are particularly difficult to translate in terms of generic stylised facts. International transport infrastructure projects involve even more decision scales, processes and stakeholders. The Oresund bridge between Denmark and Sweden is an example \cite{khan2014constructive} of such an international projects with conflicting interests between actors at different levels. The Channel tunnel also had to be planned at both the European and regional scale \cite{gibb1992channel}. The construction of the Hong-Kong-Zhuhai-Macao bridge involved geopolitical issues \cite{yang2006geopolitics}. Such projects can also be contested by the civil society as shows the example of the Lyon-Turin rail link currently under construction \cite{marincioni2009lyon}.

Therefore, the understanding of multinational transport investments and their governance remains an important aspect to account for macroscopic network dynamics. It was shown in a meta-analysis by \cite{melo2013productivity} that transport investment mostly have a positive economic effects for local growth. The One-Belt-One-Road initiative has for example already a measurable economic impact \cite{yii2018transportation}. Multinational transport investments however remain difficult to implement in practice \cite{tsamboulas1984multinational} and several trans-European projects do not provide a positive cost-benefit balance \cite{proost2014selected}. Multi-attribute theory can be used for a more multi-dimensional planning of such infrastructures \cite{tsamboulas2007tool}.

To include such governance processes at the regional, national and international scale is thus crucial to obtain more accurate macroscopic models for the co-evolution of cities and transport infrastructure. Co-evolution models including governance processes were proposed in the literature: \cite{Xie2011} describe a model of governance choice, which is spatialised into a simulation model by \cite{xie2011governance}. The aforementioned extended Luti model of \cite{raimbault2021introducing} is focused on governance level and the decision of actors to collaborate. Game theory is indeed a useful framework to model transport systems. \cite{adler2010high} model the competition between High Speed Rail and airplane. \cite{medda2007game} study public-private partnerships for transport investments. \cite{roumboutsos2008game} focus on public transport integration using game theory.

The refinement of co-evolution models between transport networks and cities at the macroscopic scale, by including more detailed processes of network investment and governance, remains to be explored. This has for example practical applications to sustainable planning at multiple scales \cite{rozenblat2018conclusion}, in particular through an understanding of the interplay between self-organisation of cities and top-down planning processes \cite{barthelemy2013self}. Our contribution in this paper is based on the model proposed by \cite{raimbault2021modeling} and the formalism to include transportation governance through game theory introduced by \cite{raimbault2021introducing}. We describe in particular a macroscopic co-evolution models which includes multiple countries and top-down governance agents making investment decisions for the development of national and international transportation links.



The rest of this paper is organised as follows: we first formally describe the simulation model and its implementation on synthetic systems of cities; we then proceed to different numerical experiments to explore its behavior; we finally discuss the implications of our results and possible developments.

\section{Model}

The macroscopic co-evolution model principally extends the model introduced by \cite{raimbault2021modeling}. It includes a geographical representation of transportation networks as introduced by \cite{raimbault2020hierarchy}. Two governance levels are involved, namely the national level and international cooperation to build cross-border infrastructures. Decisions of governance agents are simulated using the game theory module to evaluate the cooperation of agents introduced by \cite{raimbault2021introducing}.

Cities are described by their population $P_i\left(t\right)$ only and their position in space within a given country. They are linked by a physical transportation network with links described by effective distance $d_l\left(t\right)$, with a high resolution spatial description (in contrary to an abstract network description which would account only for pairwise distances between cities). The model can be understood as an iterative macro-scale Land-use Transport Interaction simulation model: at each time step, (i) spatial interaction flows between cities are updated depending on population and network distances; (ii) population of cities are evolved using these flows; (iii) network speeds are evolved depending on link flows assigned in the network.

Unconstrained spatial interaction models \cite{batty2021new} are used to estimate flows between cities, following the equation

\begin{equation}
\varphi_{ij} = \left( P_i P_j \right)^{\gamma} \cdot \exp\left(- \frac{d_{ij}}{d_G}\right)
\end{equation}

where $\gamma$ is the spatial interaction hierarchy parameter (a generalisation involving empirical data may imply a different exponent for the origin and destination cities), $d_G$ is the typical interaction distance and $d_{ij}$ is the network distance between cities.

We then assume that the growth rate of cities are proportional to cumulated interaction flows for one city, and do not include endogenous growth (Gibrat model) to focus on interaction effects. This yields

\begin{equation}
\frac{P_i \left(t + \Delta t\right) - P_i\left(t\right)}{\Delta t} \propto c_{ij} \cdot P_i\left(t\right)^{\gamma} \cdot \sum_j P_j\left(t\right)^{\gamma} \cdot \exp\left(- \frac{d_{ij}}{d_G}\right)
\end{equation}

with $c_{ij}$ a multiplier parameter equal to $1$ if cities are in the same country and $c_0 \leq 1$ (model parameter) otherwise.


Pairwise interaction flows are assigned in the network, similarly to a four-step transport model \cite{hensher2000handbook}. To simplify, we do not take into account congestion and distribute the flows following the shortest paths.

Network evolution then involves a game-theoretic approach with macroscopic agents (which in practice can be governments or planning entities). Each country has to make a choice between a national and an international investment. The following procedure is applied:
\begin{enumerate}
	\item a baseline model is computed, with a self-reinforcement of link speed according to
	\begin{equation}\label{eq:links}
	d_l\left(t \textrm{+} \Delta t\right) \textrm{=} d_l\left(t\right) \cdot \left[ 1 \textrm{+} \Delta t \cdot g_M \left( \frac{1 - \left(\frac{\varphi_l}{\varphi_0}\right)^{\gamma_N}}{1 \textrm{+} \left(\frac{\varphi_l}{\varphi_0}\right)^{\gamma_N}} \right) \right]	
	\end{equation}
	for links with $\varphi_l > \varphi_0$, where $\varphi_0$ corresponds to the $\varphi_0^{(q)}$ quantile of link flows;
	\item for each country $k$, accessibility gains with improvement of the baseline model are computed as $\Delta Z_k^{(N)}$ and $\Delta Z_k^{(I)}$, obtained respectively with national $(N)$ and international $(I)$ flows, and computed following \cite{raimbault2020unveiling}; corresponding construction costs are evaluated as $C_k^{(N)}$ and $C_k^{(I)}$ as the sum of all distance increments (linear cost);
	\item the utility matrix for the two actor game (choice between cooperation, i.e. building an international infrastructure, and non-cooperation, i.e. a national infrastructure) is given by
	
	\begin{center}
	\begin{tabular}{|c|c|c|} 
 	\hline
 	0 $|$ 1  & C & NC \\\hline
 	C & $U_i = \Delta Z_k^{(I)} - \kappa \cdot C_k^{(I)} - \frac{J}{2}$
   	& $\begin{cases} U_0 = \Delta Z_0^{(N)} - \kappa \cdot C_0^{(N)} \\U_1 = \Delta Z_1^{(N)} - \kappa \cdot C_1^{(N)} - \frac{J}{2}\end{cases}$ \\ \hline
	 NC & $\begin{cases}U_0 = \Delta Z_0^{(N)} - \kappa \cdot C_0^{(N)} - \frac{J}{2}\\U_1 = \Delta Z_1^{(N)} - \kappa \cdot C_1^{(N)}\end{cases}$
 	  & $U_i =  \Delta Z_i^{(N)} - \kappa \cdot C_i^{(N)}$ \\
 	\hline
	\end{tabular}
	\end{center}

	 what yields the mixed Nash equilibrium cooperation probabilities as
	 
	\begin{equation}
	p_i = \frac{J}{\left(Z_{1-i}^{(I)} - Z_{1-i}^{(N)}\right) - \kappa \cdot \left(C_{1-i}^{(I)} - C_{1-i}^{(N)}\right)}
	\end{equation}
	where the parameters $\kappa$ (compromise cost/accessibility benefits) and $J$ (cost of collaboration) are in practiced rescaled such that: (i) given a baseline model run, average cost absolute difference times $\kappa$ is a fixed proportion $k_0$ of average absolute accessibility difference; and (ii) collaboration cost $J$ yields a fixed probability $p_0$ computed on absolute average of the baseline run.
	\item Effective cooperation is randomly drawn for each country following these probabilities. If both countries cooperate, link distances are updated following equation~\ref{eq:links} applied with international flows, and with national flows otherwise.
\end{enumerate}



\begin{figure}
\centering
	\includegraphics[width=\linewidth]{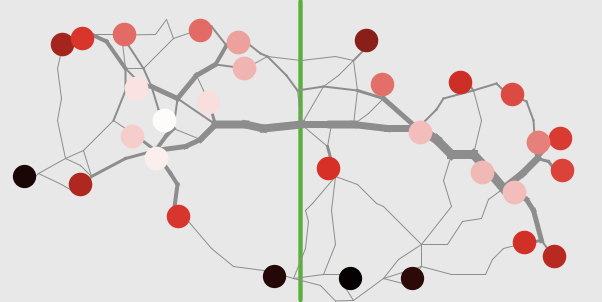}\\\hspace{0.5cm}
	\includegraphics[width=\linewidth]{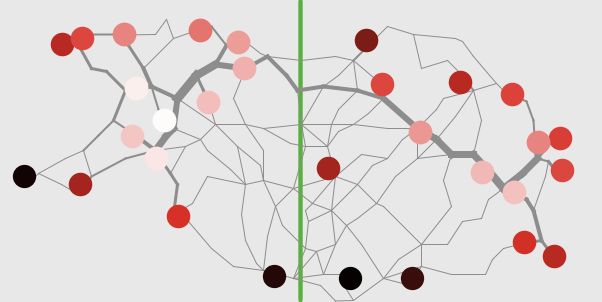}
	\caption{Example of final configuration obtained with international investments only (top) and both (bottom).\label{fig:example}}
\end{figure}

The model is setup on synthetic systems of cities \cite{raimbault2019second}. We consider a simplified configuration with two neighbour countries of the same size ($N=15$ cities, square uniform space of width 500km). Each country is setup following a Zipf law with parameters given by \cite{raimbault2020unveiling} and with a variable hierarchy $\alpha_0$, and cities are randomly assigned in space. The initial network is constructed randomly following the procedure described by \cite{raimbault2020hierarchy} with all links having a unit speed (effective distance given by the euclidian distance). An example of model setup is shown in Fig.~\ref{fig:example}.

Model indicators are built to capture cities dynamics, accessibility patterns with $\Delta P$ the average population growth and $\Delta Z$ average accessibility growth. We also study the evolution of urban hierarchy by considering $\Delta \alpha_P$ the population hierarchy difference, and $r_P$ population rank correlation. We also consider possible optimisation indicators, by computing $g$ the average effective governance level (0 being fully national and 1 fully international) and $C$ the total cost of network investment over the full model run.

Model parameters studied in experiments are $\alpha_0$ the initial hierarchy; $\gamma_G$ the spatial interaction hierarchy; $d_G$ the spatial interaction span; $c_0$ the international flows multiplier; $g_{max}$ the maximal link speed increase; $\gamma_N$ the link growth hierarchy; $\varphi_0^{(q)}$ the link growth quantile; $k_0$ relative cost of accessibility; $p_0$ effective cooperation probability; and $S$ the random seed.

\section{Results}

\subsection{Model implementation}

The model is implemented in NetLogo \cite{tisue2004netlogo}, which provides a good compromise between interactivity and efficiency, and also has fast data structures (matrix/table extensions). It is integrated into the OpenMOLE platform for model validation \cite{reuillon2013openmole}. Model code is available as open source on a git repository at \url{https://github.com/JusteRaimbault/CoevolGov}. Scripts to analyse simulation results and results are available at \url{https://github.com/JusteRaimbault/Governance}. Simulation data used in this paper are available at \url{https://doi.org/10.7910/DVN/WP4V7S}.

We use state-of-the-art model exploration methods implemented in the OpenMOLE platform to explore its behavior. We apply a Saltelli Global Sensitivity Analysis \cite{saltelli2008global}; study the role of stochasticity; and explore a simple grid sampling including the role of spatial configuration.

\subsection{Global sensitivity analysis}





First order and total sensitivity indices introduced by \cite{saltelli2008global} are obtained using 2000 models runs. Estimation results are given in Table~\ref{tab:saltelli}. We find in particular a strong effect if the spatial initial configuration through the impact of initial hierarchy $\alpha_0$ and of the random seed $S$, consistent with the important role of such factors in geographical models assessed by \cite{raimbault2019space}. We also find interaction effects between network, cities and governance.

\begin{table}
\caption{First order (F) and Total order (T) Saltelli sensitivity indices, for all indicators (rows) and parameters (columns).\label{tab:saltelli}}
\resizebox{\linewidth}{!}{
    \begin{tabular}{|l|c|c|c|c|c|c|c|c|c|c|c|c|c|c|c|c|c|c|c|c|}
\hline
 & \multicolumn{2}{|c|}{$\alpha_0$} & \multicolumn{2}{|c|}{$\gamma_G$} & \multicolumn{2}{|c|}{$d_G$} & \multicolumn{2}{|c|}{$c_0$} & \multicolumn{2}{|c|}{$g_{max}$} & \multicolumn{2}{|c|}{$\gamma_N$}  & \multicolumn{2}{|c|}{$\varphi_0^{(q)}$}  & \multicolumn{2}{|c|}{$k_0$} & \multicolumn{2}{|c|}{$p_0$} & \multicolumn{2}{|c|}{$S$} \\
 & F & T & F & T & F & T & F & T & F & T & F & T & F & T & F & T & F & T & F & T \\
 \hline
$\Delta P$ & 0.094 & 0.22 & 0.17 & 0.37 & 0.07 & 0.15 & 0.3 & 0.59 & $7\cdot 10^{-5}$ & 0.003 & -0.002 & $6.9\cdot 10^{-4}$ & -0.002 & 0.002 & -0.001 & 0.0003 & 0.002 & 0.003 & 0.02 & 0.06 \\
$\Delta Z$ & 0.05 & 0.1 & 0.02 & 0.16 & 0.52 & 0.8 & 0.02 & 0.03 & -0.006 & 0.18 & -0.006 & 0.008 & -0.008 & 0.03 & 0.0005 & 0.003 & -0.006 & 0.01 & -0.005 & 0.1 \\
$\Delta \alpha_P$ & 0.2 & 0.3 & 0.3 & 0.5 & 0.06 & 0.12 & 0.17 & 0.26 & -0.002 & 0.002 & 0.0001 & 0.0003 & -0.004 & 0.001 &-0.0007 & 0.0003 & - 0.0008 & 0.0008 & 0.01 & 0.04 \\
$r_P$ & -0.7& 0.1 & -0.1& 0.2 &-0.4 & 0.3& 0.26& 0.002 & -0.09 & 0.01 & 0.5 & 0.01 & 1.0 & 0.09 & 0.1 & 0.0002 & 0.3 & 0.001 & 0.1 & 0.07 \\
$g$ & -0.01 & 0.26 & -0.004 & 0.3 & -0.01 & 0.44 &0.03 & 0.6 & -0.03& 0.25 &-0.02 &0.2 & -0.05 & 0.3 & -0.01 & 0.2 & 0.07 & 0.7 & -0.01 & 0.5 \\
$C$ & -0.002 & 0.002 &-0.007 & 0.01 & -0.001& 0.002 & 0.002 & 0.002 & 0.06 &0.09 & 0.002 & 0.01 &0.8 & 0.9 & -0.0008 & 0.0005 &0.003 & 0.003 & 0.04& 0.04 \\\hline
\end{tabular}
}
\end{table}


\subsection{Role of stochasticity}

\begin{figure}
\centering
	\includegraphics[width=0.45\textwidth]{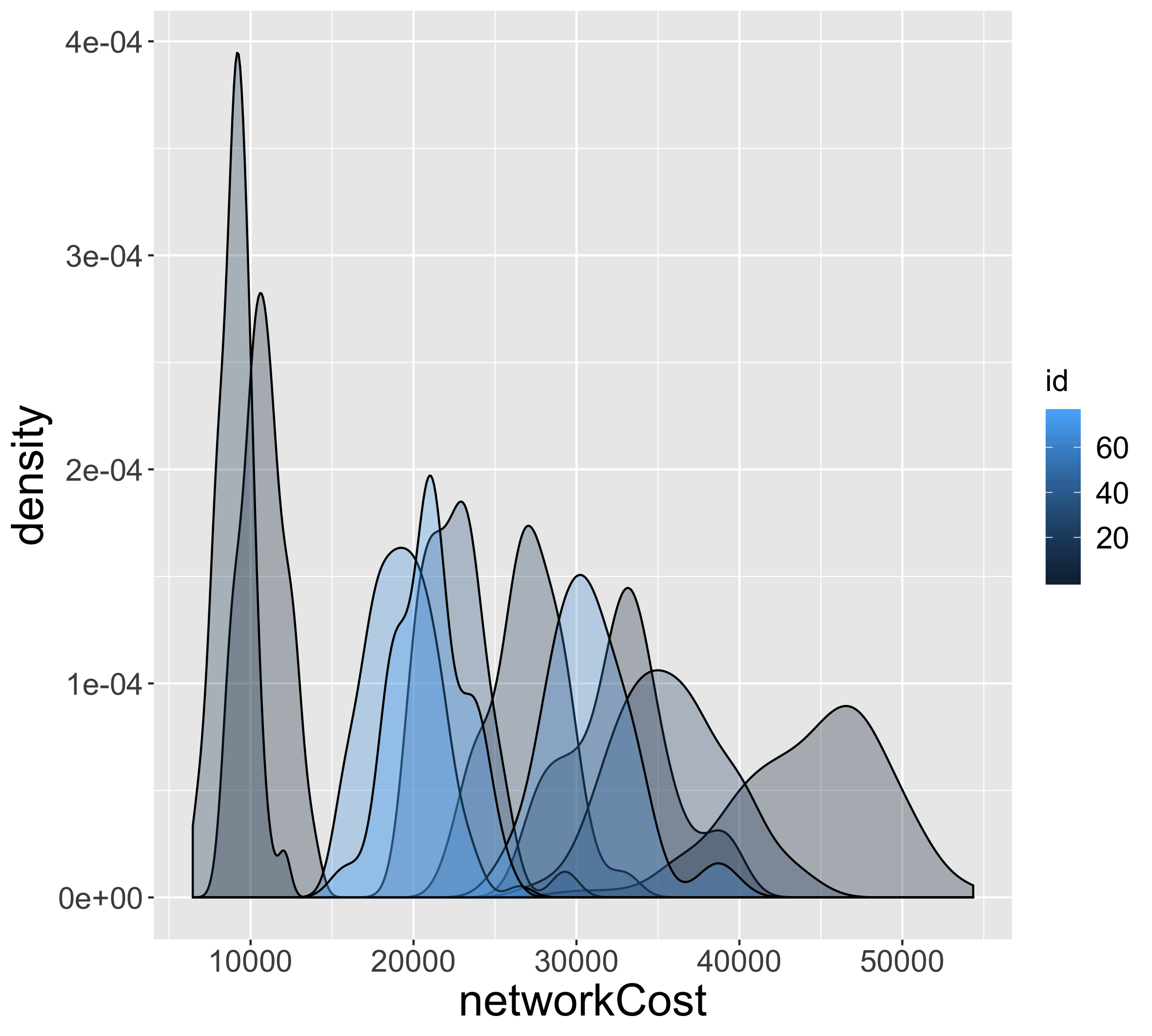}
	\includegraphics[width=0.45\textwidth]{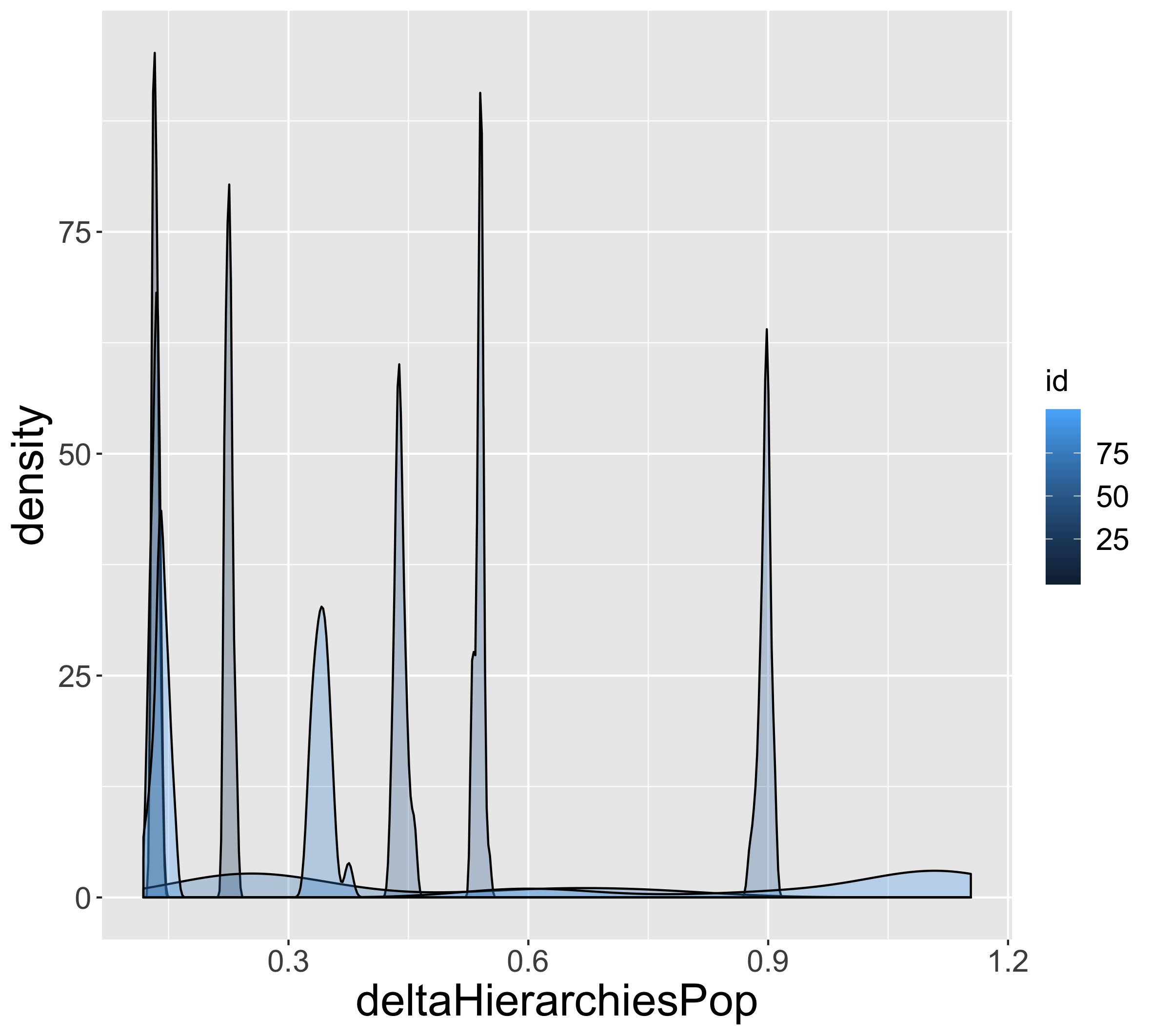}
	\caption{Histograms of network cost (left) and population hierarchy differences (right) for several points in the parameter space, obtained with 100 model replications.\label{fig:hists}}
\end{figure}

We study the role of stochasticity on the consistence of model output estimation by sampling 100 random parameters points using a Latin Hypercube Sampling, and proceed to 100 model replications for each point. We find that (i) all indicators have a high Sharpe ratio (1st quartile with a minimum of 7.7 expect governance level $g$ with a median at 1.02), implying a good convergence of indicator averages; (ii) distance between averages relative to standard deviations are also high (1st quartile higher than 2.3, except for $g$ with a median at 1.2 and $r_P$ with a median of 1.26). We show in Fig.~\ref{fig:hists} example of indicator histograms.

\subsection{Grid exploration}


\begin{figure}
\centering
	\includegraphics[width=0.48\textwidth]{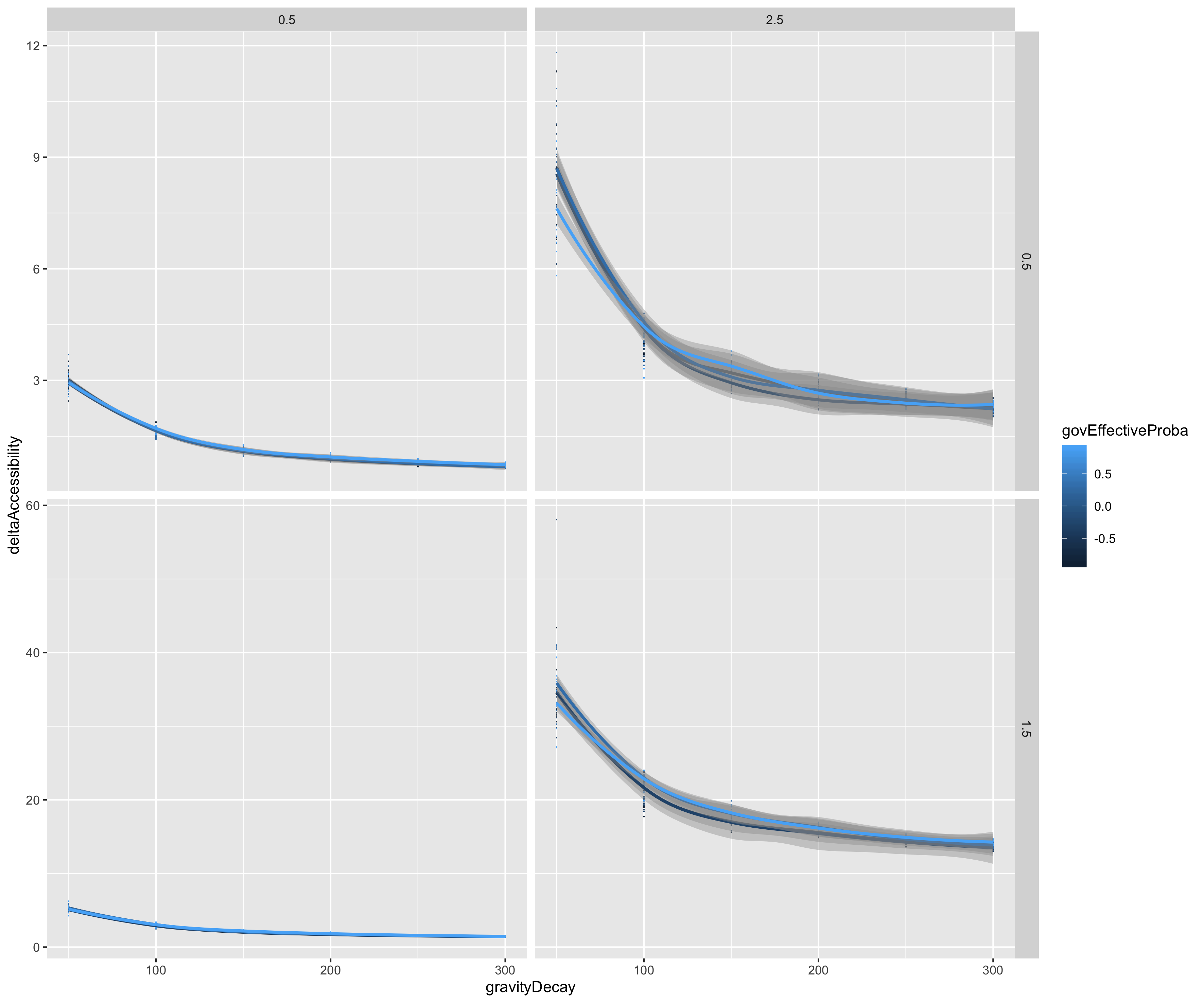}
	\includegraphics[width=0.48\textwidth]{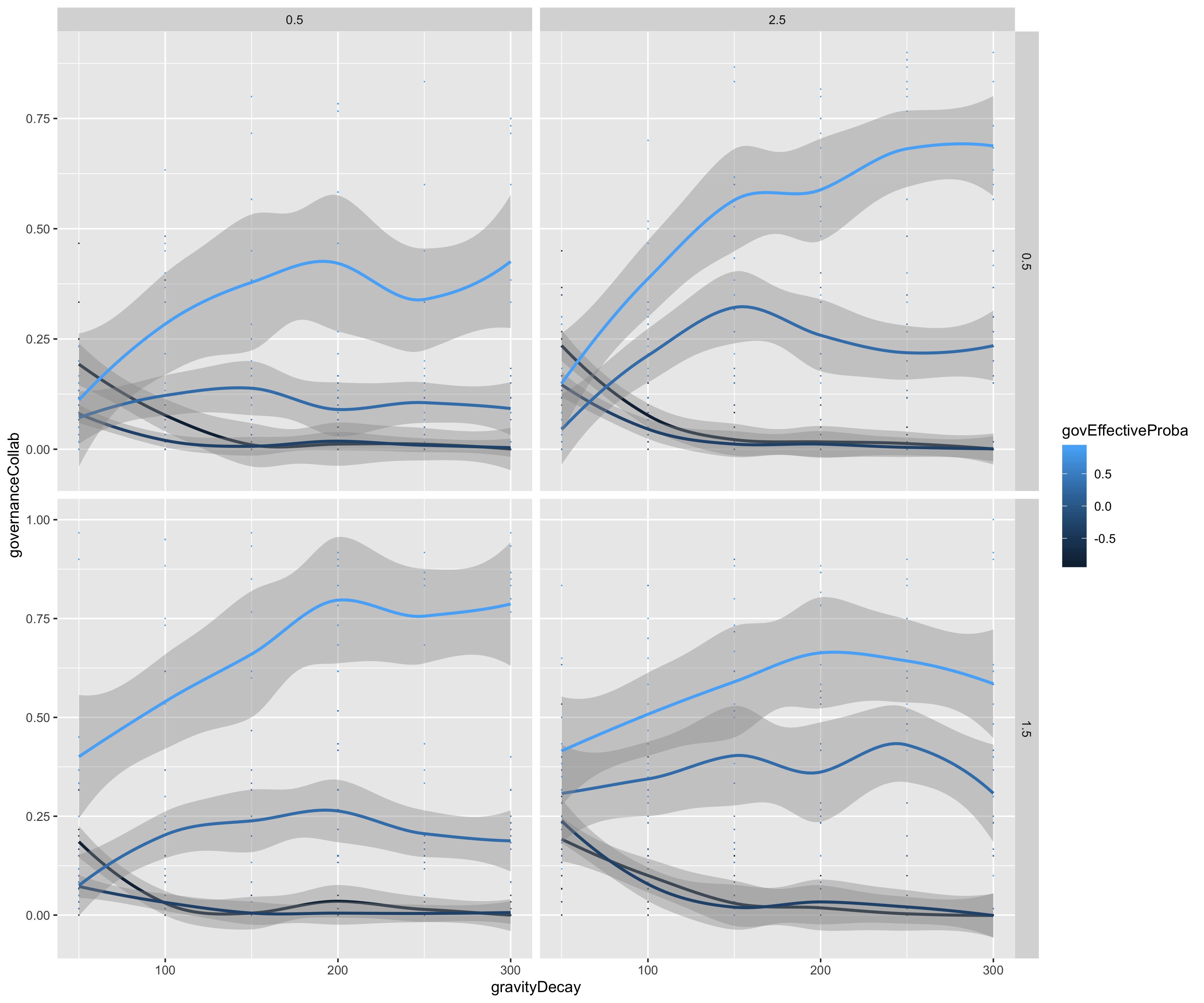}
	\caption{Accessibility gains (Left) and effective collaboration (Right) as a function of interaction span $d_G$, for different values of initial hierarchy $\alpha_0$ (rows) and interaction hierarchy $\gamma_G$ (columns).\label{fig:gridexpl}}
\end{figure}

We explore a grid of the parameter space to explore basic behavior of the model. We show in Fig.~\ref{fig:gridexpl} the evolution of accessibility gains and of effective collaboration as a function of interaction decay. We find that larger span in spatial interaction decrease relative accessibility gain, as do less hierarchical flows. However, the effective level of collaboration has no significant effect on these average accessibility gains. The effect of interaction decay on governance level is qualitatively changed by $p_0$ (switching from a decreasing to an increasing function), implying an interaction effect between geographical processes (spatial interactions) and governance processes (collaboration costs).

\subsection{Optimizing accessibility and cost}


\begin{figure}
	\centering
	\includegraphics[width=0.65\linewidth]{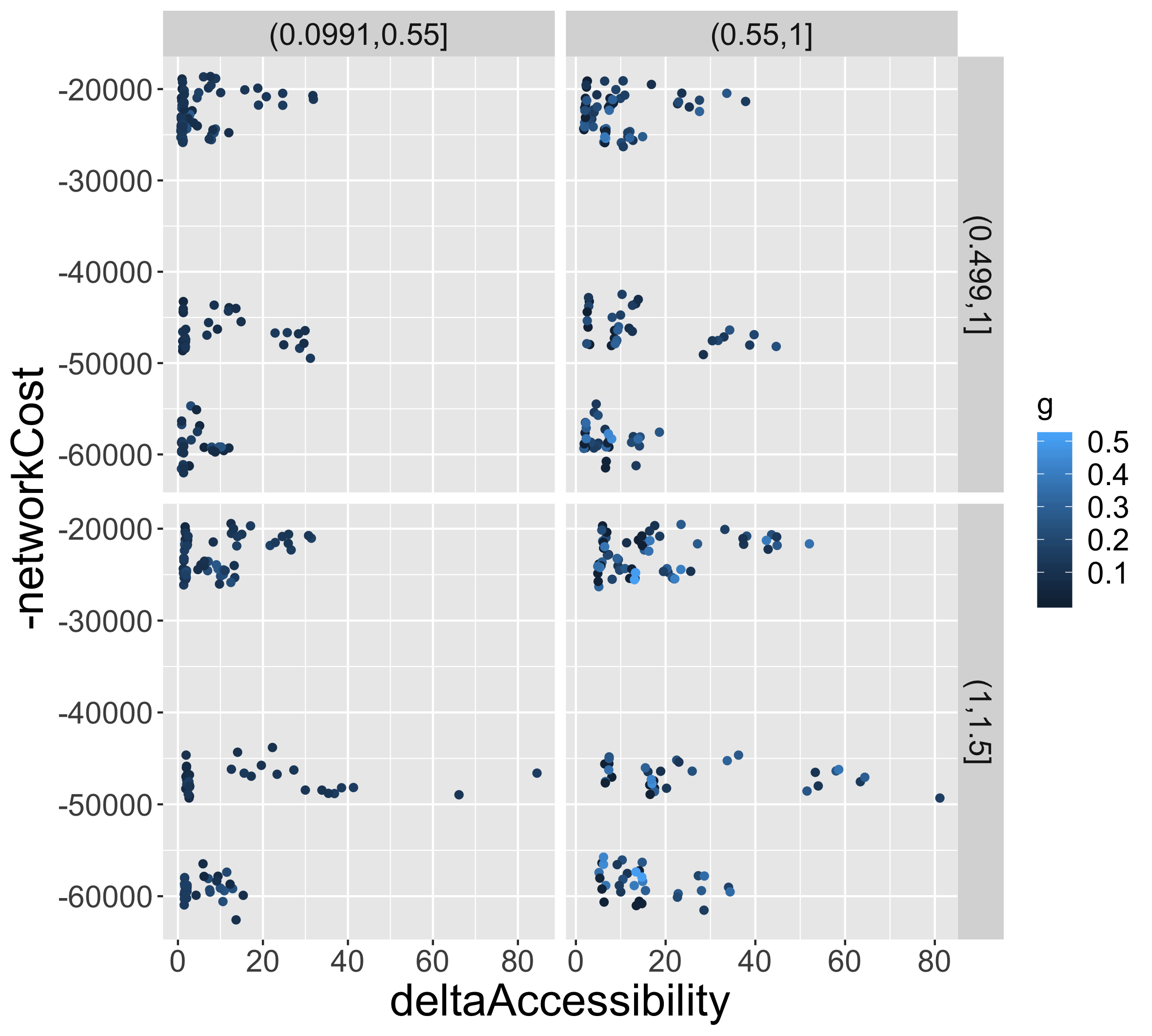}	
	\caption{Pareto fronts between accessibility gains and costs, for different levels of initial hierarchy (rows) and of international exchanges intensity ($c_0$, columns).\label{fig:pareto}}
\end{figure}

Finally, we investigate if optimisations are possible regarding network investments. Therefore, we search for compromise between accessibility gains and network costs, under the form of potential Pareto fronts. Using the previous grid experiment, we visualise the corresponding point clouds in Fig.~\ref{fig:pareto} in different initial hierarchy and international exchanges contexts. We find that Pareto-optimal points are only a small subset in each case, meaning that few optimisation compromise exist. Intermediate values of governance levels (color) are included on the optimal Pareto front when international exchanges are intense (right column). This suggests in practice a combination of local and global decisions to reach a global optimum.

\section{Discussion}


Several developments remain to be explored. Including other network assignment procedures and taking into account congestion would yield a kind of macroscopic version of Land-use Transport models. Furthermore, a more elaborated representation of decision-making and governance processes would be needed to apply the model to real situations. The coupling with \cite{raimbault2021introducing} to produce a multi-scale model is also a potential important development as such models remain rare \cite{raimbault2021strong}.

An application of this model in a realistic configuration could be useful for different reasons. First, the evaluation of transport investment scenarios or projects in a multinational, multi-stakeholder and multi-scale context is difficult to realise. The simulation model can become in that context an assessment tool. Then, more generally it could be applied to planning on long time scales for sustainable territories, by anticipating the coupled dynamics of systems of cities and infrastructure.

To conclude, this paper introduces a co-evolution model at the macroscopic scale, with an elaborated description of governance processes for the evolution of transport infrastructure. Model exploration provides a first understanding of coupled dynamics, and possibly a first step towards policy applications.





%
%
%
%
%
%
%
%


\end{document}